\documentclass[12pt]{iopart}
\usepackage{graphicx}

\begin{document}

\title{Doped Colloidal Artificial Spin Ice} 
\author{A. Lib\'{a}l$^{1}$, C.J. Olson Reichhardt$^{2}$ and C. Reichhardt$^{2}$} 
\address{
$^1$Faculty of Mathematics and Computer Science, Babes-Bolyai University, RO-400591 Cluj-Napoca, Romania \\
$^2$Theoretical Division, Los Alamos National Laboratory, Los Alamos, NM 87545, USA
}
\ead{cjrx@lanl.gov}

\begin{abstract}
We examine square and kagome artificial spin ice for colloids confined in
arrays of double-well traps.
Unlike magnetic artificial spin ices, colloidal and vortex artificial
spin ice realizations allow
creation of doping sites through double occupation of individual traps.
We find that doping square and kagome
ice geometries produces opposite effects.
For square ice, doping 
creates local excitations in the ground state configuration
that produce a local melting effect
as the temperature is raised.
In contrast,
the kagome ice ground state can absorb the doping charge without
generating non-ground-state excitations, while at elevated
temperatures the hopping of individual colloids is suppressed
near the doping sites.
These results
indicate that in the square ice, doping adds degeneracy
to the ordered ground state and creates local weak spots,
while in the kagome ice, which has a highly degenerate ground state,
doping locally decreases the degeneracy and creates local hard regions. 
\end{abstract}
\maketitle

\section{Introduction}

Artificial spin ice systems
constructed with nanomagnetic arrays \cite{1,2,3,4,5,6,7,8}, 
vortices in nanostructured superconductors \cite{9,10,11},
or soft matter systems \cite{12,13,14,15,16,17}
have attracted growing attention
as outstanding model systems in which different
types of ordered and degenerate ground states can be
realized \cite{1,8},
as well as various types
of avalanche dynamics \cite{5,18,19},
return point memory \cite{20},
and a variety of thermal effects \cite{21,22,23,25,26,27}.  
Another key feature is that many artificial spin ice systems are
constructed on size scales 
at which the microscopic degrees of freedom can be imaged directly \cite{8}.
These systems are called artificial ices since their
ground state can obey
what are called the ice rules that were
first studied in the context of a particular phase of water ice.
Here, on each bond between oxygen vertices,
two protons are localized close to
the vertex and two are localized
far from the vertex, creating what is called
the ``two-out, two-in'' rule \cite{28}. 
Since there are many ways to arrange the effective charges or spins
while satisfying the ice rule constraint,
the ground state is highly degenerate so that even
at $T = 0$ there is an extensive entropy. This 
picture has also been applied
to certain classes of atomic spin materials
with pyrochlore structures, known as spin ice systems \cite{29,30,31,32}.

\begin{figure}
\center{
\includegraphics[width=5.0in]{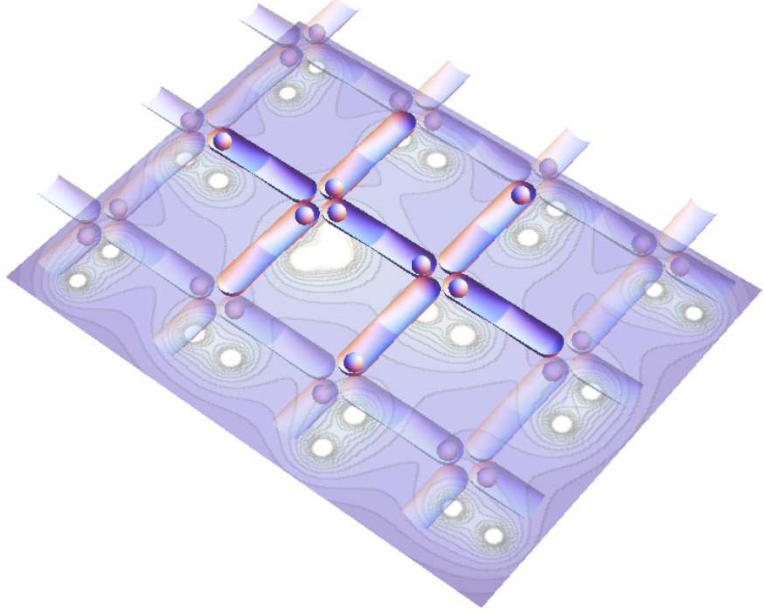}}
\caption{Illustration of a doping defect in colloidal artificial square ice.
Each trap confines   single colloid except for the center trap, which
is doped and
contains one colloid at each end.  The energy landscape created by
this configuration is plotted in the lower plane of the figure.
}
\label{fig:schematics}
\end{figure}

In the artificial spin ices, a set of geometric constraints
is imposed on the system via the arrangement of
the nanomagnetic islands \cite{8} or
arrays of double-well traps for vortices \cite{9,10,11}
or optical traps for colloids \cite{12,14}. 
Here, each trap or nanoisland plays the role of an effective macroscopic spin.
For nanomagnetic
arrays, the spin direction is defined to point in the
direction of the magnetic moment, while for the
colloidal or vortex systems with double-well traps,
the spin is defined to point toward the end of the well
that contains a particle.
The islands or traps are arranged in a square geometry
as shown in Fig.~\ref{fig:schematics} to create artificial square ice,
or in a hexagonal arrangement to realize
artificial kagome ice.
In the square ice arrangement there is an ordered ground state
in which each vertex obeys the two-in, two-out ice rule.
The lowest energy excitations in this system take the form of vertices
with three spins in or three spins out to form
a monopole with charge $-1$ or $+1$, while higher energy excitations are
vertices with four spins in or four spins out, forming monopoles
with charge  $+2$ or  $-2$ \cite{1,8}.
There are also ice rule obeying states known as biased states
that have somewhat higher energy than
the ground state, and often
pairs of monopoles
can be connected by a string of biased state vertices.
The kagome ice 
ground state is not ordered but does obey
the ice rules, which in this case are two-in, one-out or one-in, two out.
Here, monopoles consist of vertices with three spins in or
three spins out.
There are additional artificial spin ice geometries
in which the introduction of different constraints or
rules produces various types of ordered or quasiordered
ground states  \cite{33,34,35}.

It is also interesting to consider how quenched disorder can affect the ice
states.
In the magnetic spin ice systems,
disorder can arise as a dispersion in the
energy of the barrier that must be overcome
to flip an individual effective spin.
It should also be possible to 
create positional disorder in the system
by shifting the nanoislands away from their regular lattice positions.
In square ice systems, disorder can lead to the formation of domain walls,
with non-ground state vertices dividing the system into separate regions
of different ground state vertices \cite{9,6,36,37}.  Additionally,
defects on individual nanoislands can affect
the interactions between monopoles \cite{38}.
One method for characterizing the effects of disorder is by
applying a cyclic field or drive sweep to
generate hysteresis loops between two biased ground states.
Lib{\' a}l {\it et al.} constructed
hysteresis loops for colloidal artificial spin ice containing disorder in
the heights of the barriers the colloids must overcome to hop from one side
of the trap to the other \cite{20}.
They observed that
the square ice forms domain walls
that gradually coarsen during repeated cycling of the drive, increasing the
fraction of the sample that is in a ground state configuration.
In contrast, the kagome ice contained no domain walls,
and the number of non-ice rule obeying vertices at zero bias was almost
constant, indicating that
the system underwent little to no coarsening.
In the square ice the dominant mode of defect motion was
propagating domain walls,
while in the kagome ice it was
individual hopping of defects, which became pinned and
stationary rather than annihilating
\cite{20}. 

For superconducting or colloidal artificial ices,
it is possible to add disorder in the form of an effective doping by
placing doubly occupied or unoccupied traps in the sample.
A doubly occupied trap contains an
effective spin that points toward both vertices on either end of the
trap at the same time, while a doubly unoccupied (empty) trap contains
an effective spin that points away from both vertices on either end of
the trap at the same time.
In magnetic spin ice systems such configurations are not possible;
however, for water ice, additional protons can be added or 
removed at individual bonds,
allowing for similar arrangements.
In atomic spin ice systems it is possible to create what are
called stuffed spin ices by chemical 
alterations, against which
the spin ice rules have been
shown to be robust \cite{39}.  
There are also studies on diluting spin ice systems where the dilution
can generate new degrees of freedom \cite{40}.  

In this work we consider colloidal spin ice
for both square and kagome geometries that have been doped
by adding extra colloids to create doubly occupied traps.
We examine how the ices disorder under the application of thermal
fluctuations.
Colloidal systems are ideal for studying such doping
effects since large scale arrays of
optical traps can be realized and the number of colloids
per trap can be precisely tuned \cite{41,42}.
An example of the square ice geometry appears in Fig.~1, which
illustrates the local energy configuration generated by the double-well
traps, each of which captures a single colloid except for the doped site
at the center of the image which contains two colloids.
The doped site creates a geometrically necessary three-in, one-out
higher energy vertex.

\section{Simulation Method}

We perform two-dimensional Brownian dynamics simulations
of colloids in double-well traps using the same techniques applied
in previous work on colloidal spin ice systems.   
Our square ice sample contains
$N=4232$ double-well traps arranged in a
$46\times46$ lattice to form
$46\times46 = 2116$ vertices,
while our kagome ice contains
$N=4200$ double-well traps arranged in a $20\times35$ lattice
to form $20\times35\times3 = 2100$ vertices.
The simulation cell is
of size $90 a_0\times90 a_0$ for the square ice and 
$120 a_0\times121.24 a_0$ for the kagome ice system, where
$a_0$ is the simulation unit of distance, 
and it has periodic boundary conditions in both the $x$ and the $y$ directions. 
The elongated traps are $1.8a_0$ long and $0.4a_0$ wide
and are placed a center-to-center distance $a=2.0a_0$ from each other
so that they never overlap.
The overdamped equation of motion for colloid $i$ is:

\begin{equation}
\eta\frac{ d{\bf R}_{i}}{dt} = {\bf F}_{i}^{cc} + {\bf F}^{T}_{i} +
{\bf F}^{s}_{i}
\end{equation}
where the damping constant $\eta=1.0$.
The colloid-colloid interaction force has a Yukawa or screened Coulomb
form, ${\bf F}_{i}^{cc} = -F_0q^2\sum^{N}_{i\neq j}\nabla_i V(r_{ij})$
with $V(r_{ij}) = (1/r_{ij})\exp(-r_{ij}/r_0){\bf {\hat r}}_{ij}$.
Here $r_{ij}=|{\bf r}_{i} - {\bf r}_{j}|$,
${\bf {\hat r}}_{ij}=({\bf r}_{i}-{\bf r}_{j})/r_{ij}$, ${\bf r}_{i(j)}$ is the
position of colloid $i$($j$),
$F_0=Z^{*2}/(4\pi\epsilon\epsilon_0)$, $Z^*$ is the unit of charge, $\epsilon$
is the solvent dielectric constant,
$q$ is the dimensionless colloid charge, and $r_0$ is the screening length,
where $r_0=4a_0$
so that interactions extend as far as
second neighbors of each trap.
We neglect hydrodynamic interactions between colloids,
which is a reasonable assumption for charged colloids confined in 
traps that remain in the low volume fraction limit at all times.
The thermal force ${\bf F}^T$ is modeled as random Langevin kicks with the
properties $\langle{\bf F}^{T}_{i}\rangle = 0$ and
$\langle {\bf F}^{T}_i(t) {\bf F}^{T}_j(t^{\prime})\rangle
= 2\eta k_{B}T\delta_{ij}\delta(t - t^{\prime})$.
We heat the system by slowly increasing $T$ in increments of
$\delta T=0.05$ steps
from $T=0$ to $T=18$.  At each temperature we allow the system 
to equilibrate for $50,000$ simulation time steps.
As the temperature increases, the colloids begin to
hop between the two minima of the double-well traps.

\begin{figure}
  \center{
\includegraphics[width=3.5in]{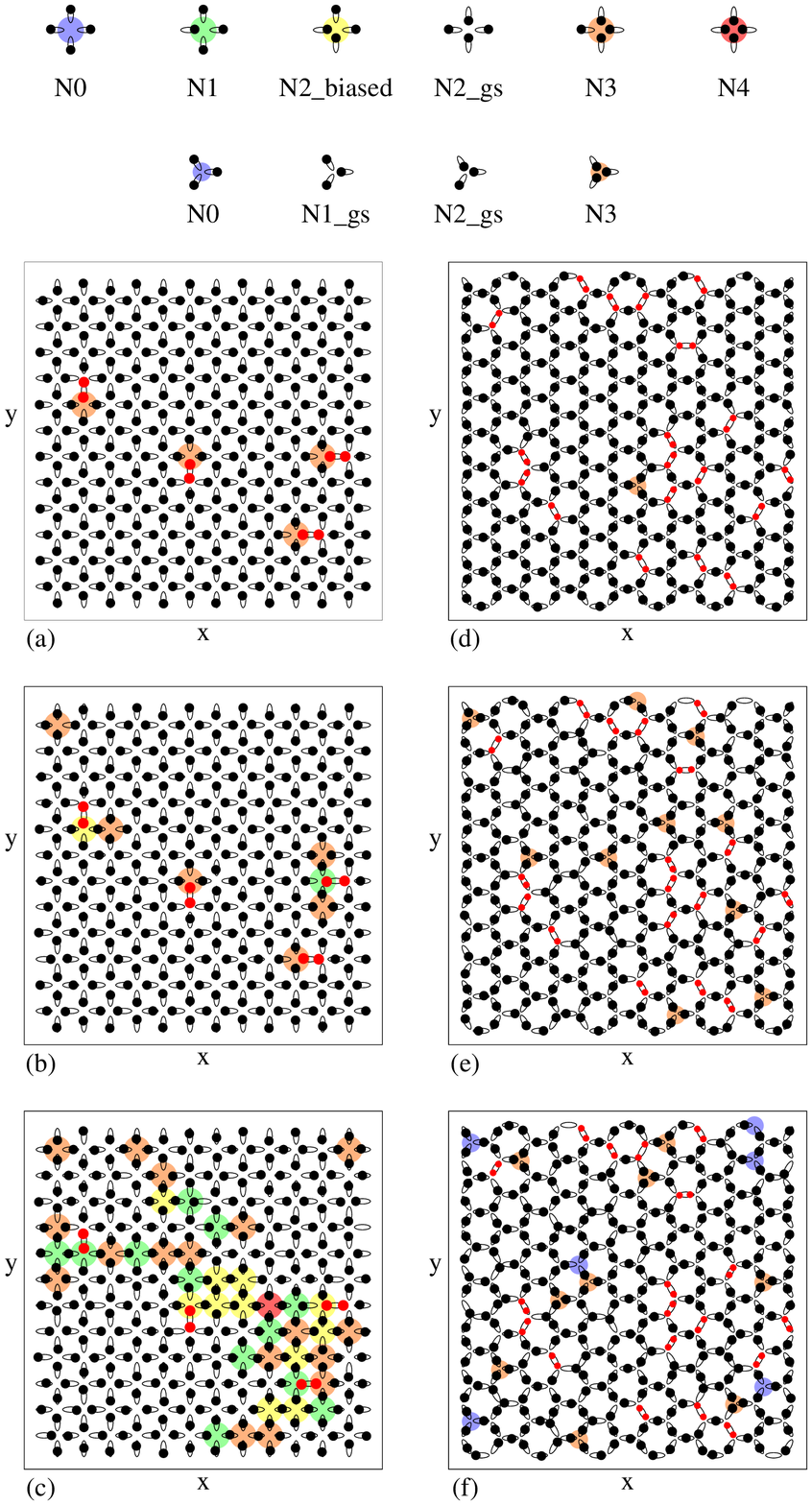}}
\caption{Images of small portions of
the square and kagome ice systems showing
the double-well traps (open ellipses),
colloids in singly-occupied traps (filled black circles), and colloids
in doubly-occupied traps (filled red circles).  
The larger colored circles indicate the vertex type as illustrated at the
top of the figure:
for the square ice, 
N0 (blue),
N1 (green),
N2$_{\rm biased}$ (yellow),
ground state N2$_{\rm gs}$ (white),
N3 (orange),
and
N4 (red);
for the kagome ice,
N0 (blue),
ground states N1$_{\rm gs}$  and N2$_{\rm gs}$ (white),
and
N3 (orange).
(a,b,c) The square ice system at a doping of $\sigma = 0.0236$. 
(a) At $T=1.0$, below melting, each doped site is screened by
an N3 monopole.
(b) At $T = 5.0$, some thermal wandering of the 
N3 sites can occur, and N1 states can form at the doped sites.
(c) At $T = 12$ the regions away from the
doped sites remain in the ground state while local
melting occurs at and near the doped sites.   
(e,f,g) The kagome ice system at a doping of $x=0.095$.
(e) At $T = 1.0$, below melting, 
the ground state
absorbs the doping charge without forming defects by
increasing the ratio of N2$_{\rm gs}$ to N1$_{\rm gs}$ vertices.
(f) At $T=12.0$, N3 monopoles form in regions away from
the doped sites.
(g) At $T=15.0$, N1 monopoles begin to appear. }
\label{fig:local_melt_schem}
\end{figure}

\begin{figure}
  \center{
  \includegraphics[width=5.0in]{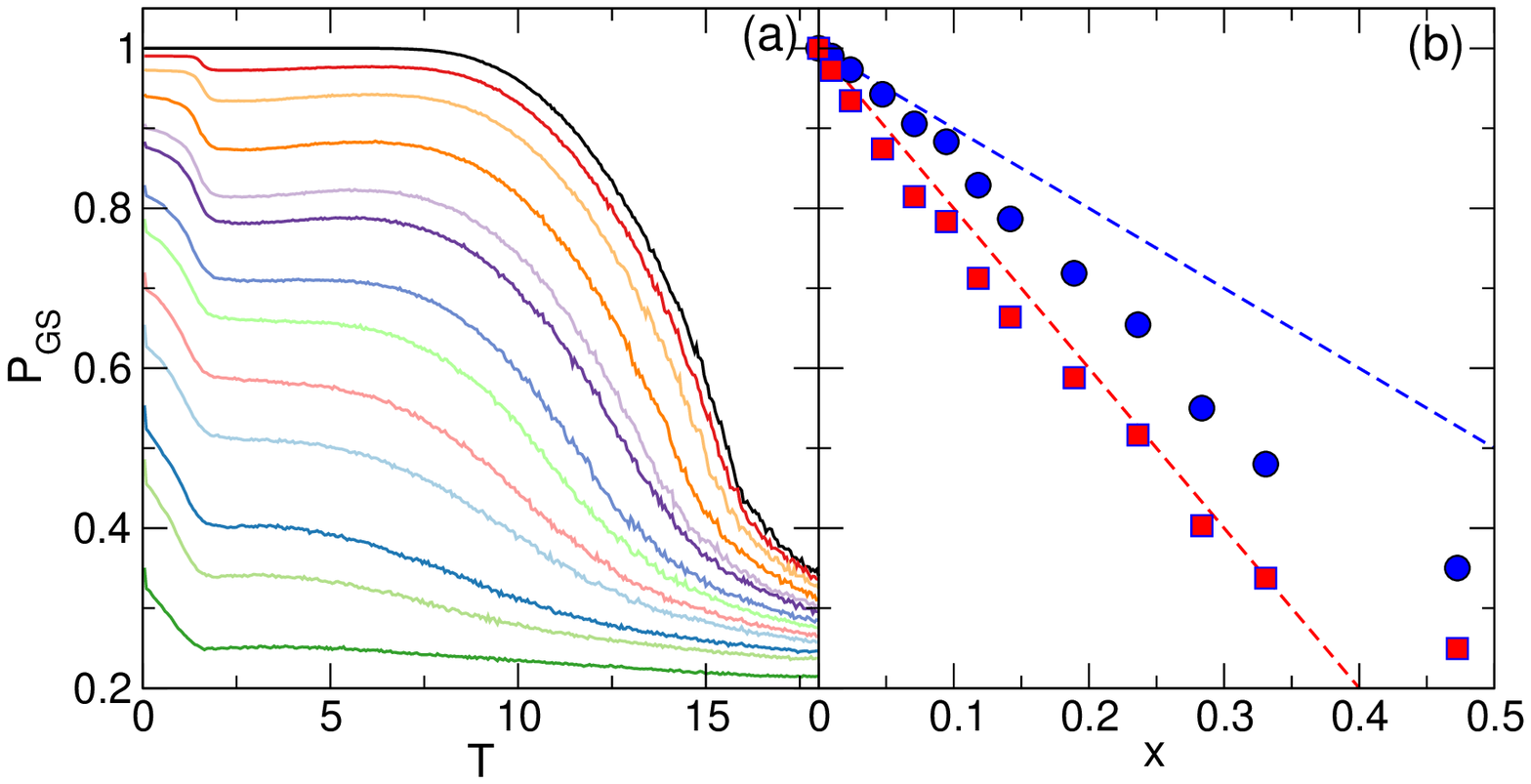}}
  \caption{ (a) The fraction of ground state vertices $P_{\rm GS}$
vs $T$ for square ice samples at doping fractions of
$x = 0$, 0.00236, 0.00472,
0.00945,
0.0709, 0.0945, 0.1182,
0.142, 0.189, 0.2363, 0.2835, $0.33$, and 0.472, from top to bottom.
A two step disordering process occurs.
(b) $P_{\rm GS}$ vs doping fraction $x$ at $T = 0$ (blue circles)
and $T = 2.0$ (red squares).
Blue dashed line: a fit to
$P_{\rm GS} = 1 - x$.
Red dashed line: a fit to $P_{\rm GS} = 1 - 2x$.   
}
\label{fig:meltT}
\end{figure}

\section{Square Ice: Local Screening and Melting }
In Fig.~\ref{fig:local_melt_schem}(a-c)
we highlight a portion of the square ice system
with a doping ratio of $x=0.0236$, where 2.36\% of the
traps are doubly occupied.  
The doped colloids are red and the different vertex types are indicated
by the colors shown at the top of the figure.
In the absence of doping, the fraction
$P_{\rm GS}$ of vertices in the
ground state at $T = 0.0$ is $1.0$, and the system becomes disordered
with $P_{\rm GS}=0.75$ near $T \approx 14$, as shown in
Fig.~\ref{fig:meltT}(a).       
We use the nomenclature
N2$_{\rm gs}$ for the ground state vertices;
N2$_{\rm biased}$ for the biased two-in, two-out states;
N3 for three-in, one-out -1 monopole vertices; 
N1 for three-out, one-in +1 monopole vertices;
N4 for a four-in -2 double monopole; and
N0 for a four-out +2 double monopole.
Figure~2(a) illustrates a low temperature
state with $T=1$,
where each doubly occupied site is screened by an adjacent N3 vertex but
there are no N1 excitations in the sample.
This is in contrast with the zero doping limit,
where for each N3 monopole excitation there must
exist a compensating N1 monopole excitation.
Such
$\pm 1$ excitation pairs can be created
through a single spin flip, while
subsequent spin flips make it possible
for the magnetic charges to move some distance away from each other through the 
lattice.
An emergent attractive force will arise between the two opposite monopoles
that can be described by a modified Coulomb law for magnetic charges,
and a Dirac string of N2$_{\rm biased}$ vertices that connects the monopoles
becomes more energetically costly as the monopoles move apart and the
string becomes longer.
In the doped system, introducing a double defect creates an
isolated N3 monopole that does not have a compensating N1 monopole.
This indicates that the square spin ice responds to the the doping
by locally screening the extra added charge through formation of
$-1$ excitations next to the double defects, as shown in 
Fig~\ref{fig:local_melt_schem}(a),
where every double defect has a screening N3 defect next to it.

In Fig.~\ref{fig:local_melt_schem}(b) we show the same
$x=0.0236$ square ice system at
$T = 5.0$.
The background ordered ground state remains frozen, but
two types of
additional excitations emerge near the doubly occupied sites.
In the first excitation, 
the screening N3 monopole starts to move away
from the doped site but remains connected to it by a string
of N2$_{\rm biased}$ vertices,
as shown in the upper left portion of Fig.~\ref{fig:local_melt_schem}(b)
where an N3 defect is separated from the doped trap
by a single N2$_{\rm biased}$ vertex.
In the second excitation,
a string of $\pm 1$ monopoles is created, as shown in the center right
portion of Fig.~\ref{fig:local_melt_schem} where the single N3 defect
has turned into a pair of N3 defects separated by an N1 defect.
In Fig.~\ref{fig:local_melt_schem}(c),
which shows the same doped square ice sample at $T = 12$,
the ground state regions away from the doubly 
occupied traps are still ordered;
however, a larger number of strings of N2$_{\rm biased}$ vertices
or N1-N3 defect strings have formed near the doped sites.
This result shows that the doping introduces 
extra topological charges that can serve as
nucleation sites for the thermal wandering of monopoles
at temperatures 
well below the bulk melting temperature.
The doping sites can be viewed as local weak spots
that induce local melting.
We also observe similar behavior at higher doping levels in the
regions that are not adjacent to the doping sites.

In Fig.~\ref{fig:meltT}(a) we plot $P_{\rm GS}$, the fraction of
vertices that are in the ground state, versus
$T$ for the square ice system from Fig.~\ref{fig:local_melt_schem}
for doping fractions $x$ ranging from
0.0
to $0.472$.
For a given temperature, $P_{\rm GS}$ decreases approximately linearly
with increasing $x$ as each doped site is screened by the formation
of a non-ground state vertex.
We observe a two-step disordering process
for finite doping,
as indicated by the drop in $P_{\rm GS}$ near $T = 1.5$,
which becomes more pronounced as $x$ increases.
At around this temperature of $T=1.5$, the N3 screening monopoles
begin to hop between neighboring sites as the individual spin degrees of
freedom within the vertex begin to flip, and as they hop,
they create additional non-ground state vertices in the form of
N2$_{\rm biased}$ or N1 monopole sites, depressing the value of $P_{\rm GS}$.
As $T$ increases further, additional thermally-activated defects appear
as the system approaches the clean melting temperature, and the resulting
decrease in $P_{\rm GS}$ begins at lower values of $T$ as the doping level $x$
increases due to interactions between vertices surrounding the randomly
placed doping sites.
In Fig.~\ref{fig:meltT}(b) we plot $P_{\rm GS}$
versus $x$ for $T=0$ and $T=2$.
At $T=0$, if each doped site created a single non-ground state vertex,
the curve would follow the blue dashed line, which is a fit to
$P_{\rm GS}  = 1 - x$.
Instead the curve drops below this value, indicating that as
the doping fraction increases and some vertices begin to interact with more
than one doped trap, more than one screening vertex can form per doping site.
At $T=2$, just above the temperature at which the screening N3 monopoles
become able to hop to an adjacent site, there is on average one additional
defected vertex generated for every screening vertex in order to permit
this hopping, as indicated by the red dashed line, which is a  
fit to $P_{\rm GS} = 1 - 2x$.

\begin{figure}
  \center{
\includegraphics[width=5.0in]{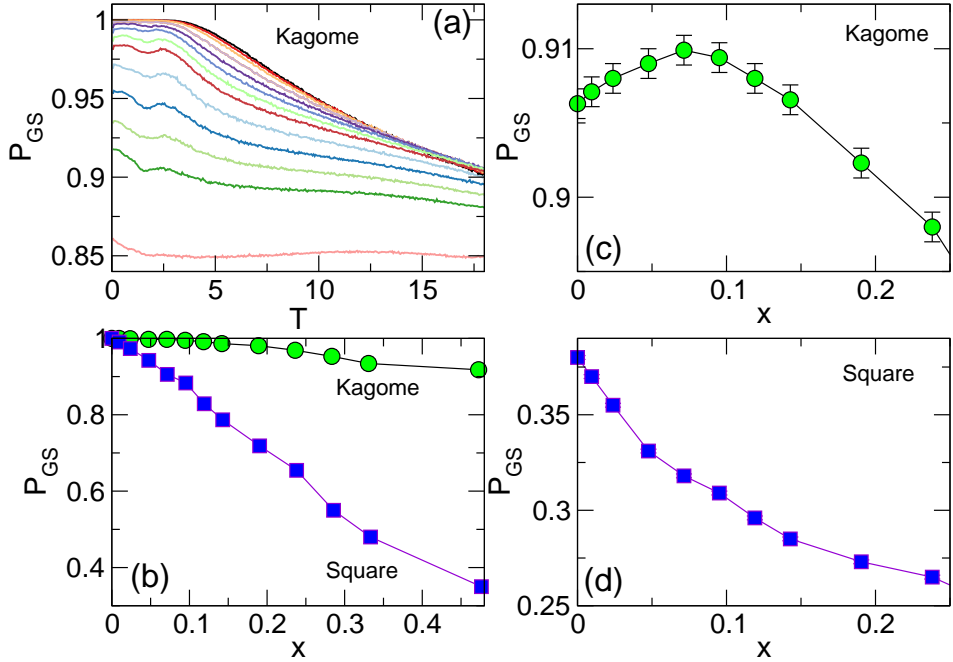}}
\caption{$P_{\rm GS}$ vs $T$ for the kagome ice system from
Fig.~\ref{fig:local_melt_schem} at
$x = 0.0$, 0.095, 0.0238, 0.0476, 0.0714, 0.0952, 0.119, 0.143, 0.19, 0.238,
0.286, 0.33, and $0.476$, from top to bottom. 
(b)  $P_{\rm GS}$ vs $x$ at $T = 0.0$ for the
kagome ice (green circles) and square ice (blue squares),
showing that doping only weakly affects
the ground state configurations of the kagome ice.
(c) $P_{\rm GS}$ vs $x$ for the kagome ice at $T = 17$, with error bars
of width 0.001. 
At this temperature, low levels of doping can reduce the defect hopping and
diminish the disorder in the ground state configuration compared to the
undoped case.
(d) $P_{\rm GS}$ vs $x$ for the square ice at $T=17$, with error bars that
are smaller than the symbols.
Addition of doping monotonically decreases the ground state
order in the system.    
}
\label{fig:kagomeT}
\end{figure}

\section{Doped Kagome Ice States}
We next consider the effects of doping on the kagome ice, as shown  
in Fig.~\ref{fig:local_melt_schem}(d,e,f)
where we highlight the vertex configurations
in a portion of a system 
with $x = 0.095$, with the red circles indicating the locations of
the doubly occupied traps.
For the kagome ice,
N1$_{\rm gs}$ and N2$_{\rm gs}$ denote the 
two-out, one-in and two-in, one-out ice rule obeying states,
respectively, 
N0 are the +1 three-out monopoles, and 
N3 are the -1 three-in monopoles.   
Figure~\ref{fig:local_melt_schem}(d) illustrates
the low temperature behavior at $T = 1.0$.  Unlike in the square ice,
addition of doping sites to the kagome ice has essentially no effect on
the ground state configuration since it is already highly degenerate.
The doubly occupied traps effectively inject
additional ``in''-pointing spins into the system.
In an undoped sample the number of N1$_{\rm gs}$ and N2$_{\rm gs}$ vertices is
roughly equal; however, when doubly occupied defects are added, the
system can absorb the extra ``in'' spins
without creating non-ground-state vertices by
increasing the fraction of N2$_{\rm gs}$ vertices in the ground state.  This
flexibility of the ground state is absent in the square ice.
As the temperature increases, we observe a continuous increase
in the number of N3 monopoles, as illustrated in
Fig.~\ref{fig:local_melt_schem}(e) at $T = 12$;      
however, there are no compensating N0 monopoles.
Formation of N3 monopoles is energetically favored
since there are an excess number of N2$_{\rm gs}$ vertices in the doped
ground state.
At $T = 15$, shown in Fig.~\ref{fig:local_melt_schem}(f),
the thermal fluctuations are large enough to create both N0 and N3 vertices.  
We find that the doped traps in the kagome ice generally do not act
as nucleation sites for additional defects.
For higher $T$, the colloids begin to escape from the double-well
traps, which sets an upper limit on the range of temperature we can study. 

To better understand why the doping has little effect on the kagome
ground state,
in Fig.~\ref{fig:kagomeT}(a) we plot   
$P_{\rm GS}$ versus $T$ for kagome samples with $x$ ranging from $x=0$ to
$x=0.476$.
In Fig.~\ref{fig:kagomeT}(b), the $P_{\rm GS}$ versus
$x$ curves for the kagome and square ices at $T=0$ indicate that
the ground state configuration is lost much more rapidly with increasing
doping in the square ice than in the kagome ice.
There is a small dip in $P_{\rm GS}$ in Fig.~\ref{fig:kagomeT}(a) near
$T=2.0$ for intermediate doping, followed by a local maximum in
$P_{\rm GS}$ at $T=5$.
By condensing some of the ``in''-pointing spins into
N3 vertices, the ratio of N1$_{\rm gs}$ to N2$_{\rm gs}$
ground state vertices
can be shifted back towards its unbiased level.
For $T<2.0$, the decrease in $P_{\rm GS}$ with increasing $T$ in the
doped samples occurs as the energy barrier to condensation of
N3 vertices out of the N2$_{\rm gs}$-biased ground state is overcome
thermally at locations near doping sites where this
barrier is suppressed, leading to the formation of larger numbers
of N3 vertices
as the temperature rises.
This process halts at $T=2.0$, the melting temperature of N3 vertices
in an undoped system.  For $2.0<T<2.5$, the N3 vertices melt back into
N2$_{\rm gs}$ vertices and $P_{\rm GS}$ increases with increasing $T$.
The ground state melting temperature is $T=2.5$, as shown by the black
line in Fig.~\ref{fig:kagomeT}(a) for the undoped sample, so for
$T=2.5$ and above, $P_{\rm GS}$ decreases with increasing $T$ as
thermally excited defects emerge throughout the sample.

For $T>14$ and low doping, some of the doped states develop higher
ground state order than the undoped sample.
This is more clearly illustrated in
Fig.~\ref{fig:kagomeT}(c), where we plot
$P_{\rm GS}$ versus $x$ at $T = 17$. 
Here $P_{\rm GS}$ initially increases with increasing doping
before reaching a local maximum at $x = 0.075$ and
then decreasing as the doping is further increased.
This shows that in certain cases the doping can suppress
the thermally induced creation of N0 and N3 defects by locally
breaking the degeneracy of the kagome ground state.
For comparison, in Fig.~\ref{fig:kagomeT}(d) we plot
$P_{\rm GS}$ versus $x$ at $T=17$ for the square ice, where we find a
monotonic decrease in $P_{\rm GS}$ with increasing doping. 

\begin{figure}
  \center{
\includegraphics[width=5.0in]{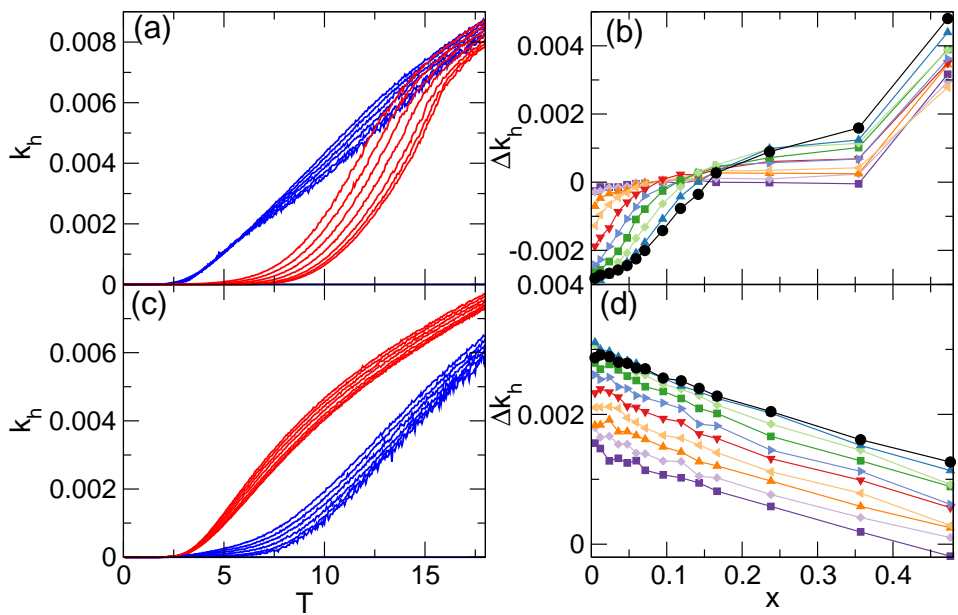}}
\caption{(a) Hopping rate $k_h$ vs $T$ for the square ice system.
Blue curves: $k_h^c$ for vertices adjacent to doubly-occupied traps;
red curves: $k_h^f$ for vertices not adjacent to doubly-occupied traps.
The curves are plotted for
$x = 0$, 0.00945, 0.0236, 0.0472, 0.0709, 0.0945,
and $0.1181$, from the bottom red (blue) curve to the top red (blue) curve.
The hopping rate is enhanced near the doping sites in the square ice.
(b) The difference in the hopping rates
$\Delta k_h=k_h^f-k_h^c$
vs $x$ for
$T=8$ (black circles),
$T=9$ (dark blue up triangles),
$T=10$ (light green diamonds),
$T=11$ (dark green squares),
$T=12$ (light blue right triangles),
$T=13$ (red down triangles),
$T=14$ (light orange left triangles),
$T=15$ (dark orange up triangles),
$T=16$ (light purple diamonds),
and $T=17$ (dark purple squares).
For small values of $x$,
$\Delta k_h$ is negative since the hopping rate is
highest close to the doping sites, and it approaches zero as the doping level
or temperature is increased.
(c)
$k_h^c$ (blue) and $k_h^f$ (red) vs $T$ for the kagome ice
at $x = 0$, 0.0094, 0.0238, 0.0476, 0.07142, 0.0952, and  0.119
from the bottom red (blue) curve to the top red (blue) curve.
Here the trend in the hopping rate is reversed,
with the lowest hopping rates close to the doping sites.
(d) $\delta k_h$ vs $x$ for the kagome ice for $T=8$ to $17$ from top to
bottom, with the same symbols as in panel (b).
}
\label{fig:hop}
\end{figure}

\begin{figure}
  \center{
\includegraphics[width=5.0in]{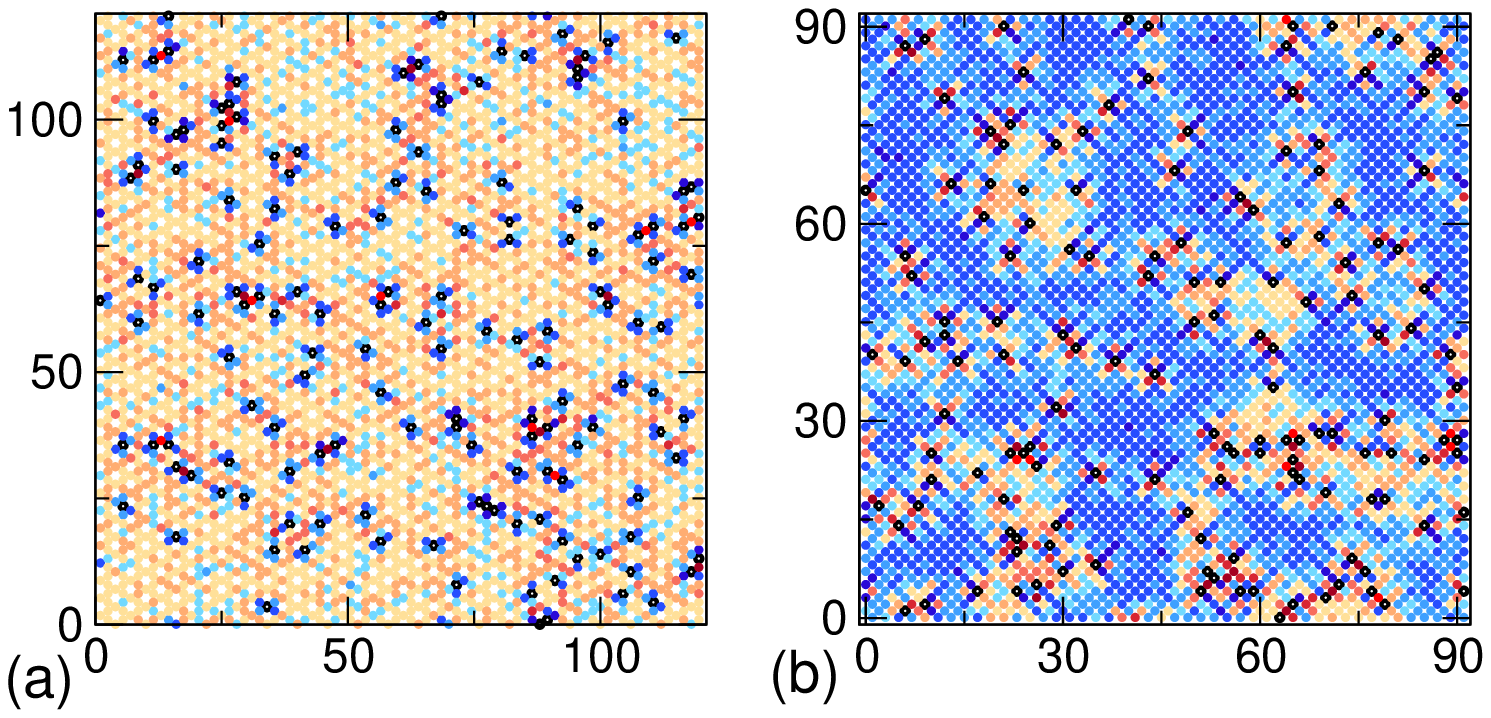}}
\caption{ 
The spatial distribution of the thermally induced hopping in the entire
sample
at $T = 12.0$.  Black circles indicate the doped traps.
Hopping rates are indicated on a color scale where red sites
have high $k_h$ and blue sites have low $k_h$.
(a) In the kagome ice at $x = 0.07142$, traps near the doping sites
have suppressed hopping rates.
(b) In the square ice system at $x = 0.0709$, traps near the doping sites
have increased hopping rates.
}
\label{fig:hopimage} 
\end{figure}

\section{Dynamics} 

To further quantify the different effects of doping
in the square and kagome ices, we examine the
hopping rate $k_h$ at which the colloids jump from one well to the other
inside the double-well traps, in units of inverse simulation time steps.
We average $k_h$ over two colloid populations:  $k_h^c$ is the average hopping
rate for all colloids in traps that are next to
doped, doubly-occupied traps, while $k_h^f$ is the average hopping rate
for the colloids in traps that are not next to
doping sites.  Here, a trap is defined to be next to a doping site if
it is part of a vertex that includes the doped trap.
In Fig.~\ref{fig:hop}(a) we plot $k_h^c$ and $k_h^f$ versus $T$ for
square ice samples with $x$ ranging from $x=0$ to $x=0.1181$.
For finite doping, $k_h^c$ for the traps close to
doping sites increases from zero near $T=2$, while $k_h^f$ for the traps
that are away from doping sites does not begin to increase until
$T=5$ to 7.
The temperature at which the initial upturn of $k_h^f$ occurs
decreases with increasing $x$, caused when the hopping barrier in
an undoped site is depressed by proximity to two or more doped sites,
an arrangement that becomes more common as $x$ increases.
This effect appears more clearly 
in Fig.~\ref{fig:hopimage}(b) where we plot the spatial distribution of
hopping rates in a
square ice system with $x = 0.0709$ at $T = 12$.
Here the largest hopping rates occur for traps that are immediately
adjacent to two or more doping sites.
In Fig.~\ref{fig:hop}(b) we plot the difference in the hopping rates
$\Delta k_h=k_h^f-k_h^c$ as a function of
$x$ for square ice samples with $T =  8$ to 17.
For low doping $x < 0.2$, the magnitude of $\Delta k_h$
is largest at low temperatures
and gradually diminishes to zero as the temperature increases.
For $x > 0.2$ we find a reversal in the
relative hopping rates, where the hopping switches from being more rapid
close to doping sites at low $x$ to being more rapid away from doping sites
at higher $x$.
This reflects a transition to a regime where the density of doped sites
becomes high enough to strongly constrain the colloid positions in
many undoped sites due to random clustering of the doped sites,
so that traps close to doped sites have a suppressed rate
of thermally induced hopping.

In Fig.~\ref{fig:hop}(c) we plot the hopping rates
$k_h^c$ and $k_h^f$ versus $T$
for the kagome ice at doping levels
$x = 0.0095$ to $0.119$.   
Here we find the opposite behavior from the square ice, with
colloids in traps next to doping sites having
a significantly lower hopping rate than colloids
in traps that are away from doping sites.
Figure~\ref{fig:hop}(d) shows $\Delta k_h$ is
largest at the lowest temperatures and monotonically decreases
both with increasing $T$ and with increasing $x$.
In Fig.~\ref{fig:hopimage}(a)
we illustrate the spatial distribution of the hopping rates
in the kagome ice
for $x=0.0714$ at $T = 12$, where the lowest hopping rates
appear adjacent to the doped sites.
The clustering effect observed for the hopping rate in the
square ice system, where there is a change in sign of
$\Delta k_h$ as the doping level increases, is absent in the
kagome ice.
These results show that doping has opposite effects
on square and kagome ices.
Since the square ice has an ordered ground state, the doping
adds local frustration which serves as weak spots at which
hopping can more readily occur.
In the kagome ice, the ground state is not
ordered, and doping instead lifts the local degeneracy
to create ``hard'' spots at which the hopping rate is suppressed.

\section{Summary}

In summary, we have examined the effect of doping
on square and kagome artificial spin ice systems constructed from
colloids in double well traps.
Doping is introduced by adding an extra colloid to a single trap
to create an effective spin that is pointing both in and out of
the corresponding vertex.
For the square ice, doping
creates additional monopole excitations that serve to screen
the doped site from the bulk.  As the temperature increases, these
screening monopoles begin to
move away from the doping sites by generating additional monopoles or
a string of biased ground state vertices.
The doping sites create local weak spots at which
local thermal disordering can readily occur
as the temperature is increased.
For kagome ice, we find the opposite effect, with no additional
monopoles created by doping since a shift in the population of ground
state vertices serves to absorb the extra charge introduced by doping,
while there is a reduction of the hopping rate of
colloids in traps adjacent to the doped sites.
The doping can cause a small increase in the fraction 
of vertices in the ground state configuration at finite temperatures
in the kagome ice due to the suppression of the
hopping, while for the square ice the doping always
decreases the fraction of vertices in the ground state.  

\section{Acknowledgements}
We thank C. Nisoli for useful discussions.
This work was carried out under the auspices of the 
NNSA of the 
U.S. DoE
at 
LANL
under Contract No.
DE-AC52-06NA25396.
The work of AL was supported by a grant of the Romanian National Authority for Scientific Research, CNCS-UEFISCDI, project number PN-II-RU-TE-2011-3-0114.

\end{document}